\documentclass[aps,showpacs,preprint]{revtex4}
\usepackage{psfig,rotating}
\begin{document}
\title{Quantum state engineering assisted by entanglement}
\author{Matteo G. A. Paris}
\affiliation{Quantum Optics \& Information Group,
INFM Unit\`a di Pavia, Italia}
\author{Mary Cola and Rodolfo Bonifacio}
\affiliation{Dipartimento di Fisica and Unit\`a
INFM, Universit\'a di Milano, Italia.}
\date{\today}
\begin{abstract}
We suggest a general scheme for quantum state engineering based on
conditional measurements carried out on entangled twin-beam of
radiation. Realistic detection schemes such as {\sc on/off}
photodetection, homodyne detection and joint measurement of
two-mode quadratures are analyzed in details. Imperfections of the
apparatuses, such as nonunit quantum efficiency and finite
resolution, are taken into account. We show that conditional {\sc
on/off} photodetection provides a reliable scheme to verify
nonclassicality, whereas conditional homodyning represents a
tunable and robust source of squeezed light. We also describe
optical teleportation as a conditional measurement, and evaluate
the degrading effects of finite amount of entanglement,
decoherence due to losses, and nonunit quantum efficiency. 
\end{abstract}
\pacs{03.67.Mn,42.50.Dv}
\maketitle
\section{Introduction}\label{s:intro}
Quantum state engineering of radiation field plays a major role
in several fundamental tests of quantum mechanics \cite{ts0}, as
well as in applications such high precision measurements and high
capacity communication channels \cite{qcm}.  Generation of
nonclassical light generally involves active devices and
nonlinear optical media, which couple two or more modes of the
field through the nonlinear susceptibility of the matter. Since
the nonlinear susceptibilities are small, the effective
implementation of nonlinear interactions is experimentally
challenging, and the resulting processes are generally
characterized by a low rate of success, {\em i.e.} by a low
efficiency.
\par
In quantum mechanics, the reduction postulate provides an
alternative {\em intrinsic} mechanism to achieve {\em effective}
nonlinear dynamics.  In fact, if a measurement is performed on a
portion of a composite entangled system, {\em e.g.} the bipartite
entangled systems made of two modes of radiation, the other
component is conditionally {\em reduced} according to the outcome
of the measurement \cite{vn}. The resulting dynamics is highly
nonlinear, and may produce quantum states that cannot be
generated by currently achievable nonlinear processes. The
efficiency of the process, {\em i.e.} the rate of success in
getting a certain state, is equal to the probability of obtaining
a certain outcome from the measurement.  This is usually higher
than nonlinear efficiency, thus making conditional schemes
possibly convenient even when a corresponding Hamiltonian process
exists.
\par
The nonlinear dynamics induced by conditional measurements has
been analyzed for a large variety of tasks
\cite{dar,wel,pgb,cat1,cat2,cat3,ff,zlc,uni,opq,cla,sab,pla,bno,koz},
among which we mention photon adding and subtracting schemes
\cite{wel}, optical state truncation of coherent states
\cite{pgb}, generation of cat-like states \cite{cat1,cat2,cat3},
state filtering by active cavities \cite{ff,zlc}, synthesis of
arbitrary unitaries \cite{uni}, and generation of optical qubit
by conditional interferometry \cite{opq}.
\par
In this paper we analyze in details the use of conditional
measurements on entangled twin-beam (TWB) to engineer quantum
states, {\em i.e.} to produce, manipulate, and transmit
nonclassical light. In particular, we will focus our attention on
realistic measurement schemes, feasible with current technology,
and will take into account imperfections of the apparatuses such
as detection quantum efficiency and finite resolution.
\par
The reason to choose TWB as {\em entangled resource} for
conditional measurements is twofold. On one hand, TWBs are the
natural generalization to continuous variable (CV) systems of Bell
states, {\em i.e.} maximally entangled states for qubit systems.
On the other hand, and more important, TWBs are the only CV
entangled states that can be reliably produced with current
technology, either by parametric downconversion of the vacuum in a
nondegenerate parametric amplifier \cite{kum0}, or by mixing two
squeezed vacua from a couple of degenerate parametric amplifiers
in a balanced beam splitter \cite{kim,joi}. Overall, our main goal
is to establish the current state of art for conditional
engineering of CV quantum states assisted by entanglement.
\par
The first kind of measurement we analyze is {\sc on/off}
photodetection.  As a matter of fact, though recent proposals are
encouraging \cite{kwt}, the discrimination of, say, $n$ photons
from $n+1$ photons in the quantum regime is still experimentally
challenging. Therefore, we are led to consider the action of
realistic avalanche {\sc on/off} photodetectors, i.e. detectors
which have no output when no photon is detected and a fixed
output when one or more photons are detected. Our analysis shows
that {\sc on/off} photodetection on TWB provides the generation
of conditional {\em nonclassical mixtures}, which are not
destroyed by decoherence induced by noise and permits a robust
test of the quantum nature of light. The second apparatus is
homodyne detection, whose action on TWB represents a tunable
source of squeezed light, with high conditional probability and
robustness to experimental imperfections, such nonunit quantum
efficiency and finite resolution.  The third kind of measurement
we are going to consider is the joint measurement of the sum- and
difference-quadratures of two modes, corresponding to the
measurement of the real and the imaginary parts of the complex
photocurrent $Z=a+b^\dag$, $a$ and $b$ being two modes of the
field.  Such a measurement is realized by generalized heterodyne
detection if the two modes have different frequencies, and by
multiport homodyne detection if they have the same frequency. In
our case one of the two modes is a beam of the TWB, whereas the
second mode is excited in a given reference state, usually
referred to as the probe of the measurement. As we will see, this
approach allows to describe CV quantum teleportation as a
conditional measurement, and to easily evaluate the degrading
effects of finite amount of entanglement, decoherence due to
losses, and imperfect detection.
\par
The paper is structured as follows. In Section \ref{s:main} we
establish notation and describe the general measurement scheme we
are going to consider. In Section \ref{s:details} we consider the
three above detection schemes as conditional measurements to
engineer nonclassical states. In Section \ref{s:wig} we show how
to evaluate detection probabilities and conditional states using
Wigner functions. This approach allows us to analyze several degrading
effects in CV teleportation, and to show the equivalence of noisy
teleportation to a Gaussian noisy channel.  Section \ref{s:outro} closes the
paper with some concluding remarks.
\section{Conditional quantum state engineering}
\label{s:main}
\begin{figure}[h]
\psfig{file=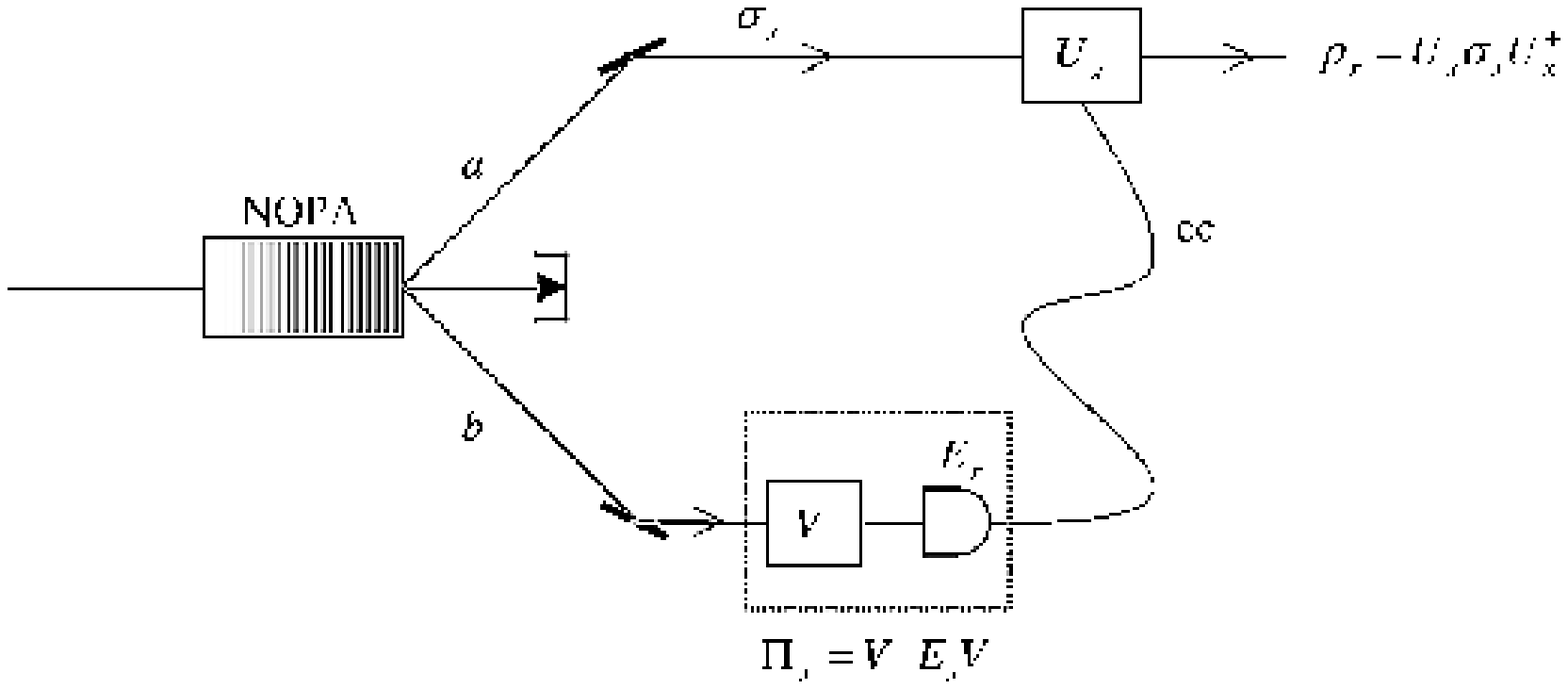,width=8cm}
\caption{Scheme for quantum state engineering assisted by entanglement.
At first, a twin-beam of the modes $a$ and $b$ is produced by spontaneous
downconversion in a nondegerate parametric optical amplifier. Then, mode
$b$ is (possibly) subjected to the unitary transformation $V$ and then
revealed by a measurement apparatus described by the POVM $E_x$. Overall, the
quantum operation on the mode $b$ is described by the POVM $\Pi_x=V^\dag E_x
V$. The conditional state of mode $a$ is given by $\varrho_x$: this state may
be further modified by a unitary transformation $U_x$ depending on the outcome
of the measurement, whose value may be sent to the receiver location by classical
communication. We always take $V=I$ (no transformation before the measurement),
and consider three kind of measurements: {\sc on/off} photodetection, homodyne
detection and joint measurement of two-mode quadratures by multiport
homodyne or heterodyne detection.
In the case of {\sc on/off} photodetection and homodyne detection we do not
consider further transformation ({\em i.e.} $U_x=I$), whereas for heterodyne
detection this is a displacement operator $D(\alpha)$, with amplitude equal to the
result of heterodyne detection.}\label{f:scheme}
\end{figure}

The general measurement scheme we are going to consider is schematically
depicted in Fig. \ref{f:scheme}. The first stage consists of
a non-degenerate optical parametric amplifier (NOPA) obtained by a
$\chi^{(2)}$ nonlinear optical crystal cut either for type I or type II
phase-matching. In the parametric approximation ({\em i.e.} pump remaining
Poissonian during the evolution \cite{para}) the crystal couples two modes
of the radiation field according to the effective Hamiltonian
\begin{eqnarray}
 H_\kappa =\kappa (a ^\dag b ^\dag + a b) \;,  \label{nopa}
\end{eqnarray}
where $\kappa$ represents the effective nonlinear coupling, and $a$ and $b$
denote modes with wave vectors satisfying the phase-matching condition $
\vec k_a + \vec k_b =\vec k_p$, $\vec k_p$ being the wave vector of the
pump. For vacuum input we have parametric downconversion, with the output
given by the so-called {\em twin beam} state of radiation
\begin{eqnarray}
|\lambda \rangle\rangle = \sqrt{1-|\lambda |^2}\: \sum_{p=0}^\infty \:
\lambda^p \: |pp\rangle\rangle \quad
|pp\rangle\rangle = |p\rangle_a \otimes |p\rangle_b
\label{twbdef}\;
\end{eqnarray}
where $\lambda = \tanh|\kappa|\tau$ and $\tau$ represents an
effective interaction time. The TWB $|\lambda\rangle\rangle$ is an
entangled state living in the bipartite Hilbert space ${\cal
H}_a\otimes {\cal H}_b$, where ${\cal H}_j$, $j=a,b$, are the Fock
space of the two modes respectively. TWBs are pure states and thus
the entanglement can be quantified by the excess Von-Neumann
entropy \cite{bp91,lind,bp89,ve97}. The entropy of a two-mode
state $\varrho$ is defined as $S[\varrho] = -
\hbox{Tr}\left\{\varrho \log\varrho\right\}$ whereas the entropies
of the two modes $a$ and $b$ are given by $S[\varrho_j] =
-\hbox{Tr}_j\left\{\varrho_j \log\varrho_j\right\}$, $j=a,b$, with
$\varrho_a = \hbox{Tr}_b\left\{\varrho\right\}$ and $\varrho_b =
\hbox{Tr}_a\left\{\varrho\right\}$ denoting partial traces.
The degree of entanglement of the state $\varrho$, in terms 
of the average number of photons of the TWB $N=2 
\lambda^2/(1-\lambda^2)$, is given by
\begin{eqnarray}
\Delta S &=& \frac12 (S[\varrho_a] + S[\varrho_b] -
S[\varrho]) \nonumber \\ &=& \log (1+\frac{N}{2}) + 
\frac{N}2 \log (1+\frac{2}N)
\label{DeltaS}\;.
\end{eqnarray}
Notice that for pure states $\Delta S$ represents the unique measure of
entanglement \cite{po97}, and that TWBs are maximally entangled states for a
given (average) number of photons. The degree of entanglement is a
monotonically increasing function of $N$.  
\par
A measurement performed on one of the two modes {\em reduces} the
other one according to the projection postulate. Each possible
outcome $x$ occurs with probability $P_x$, and corresponds to a
conditional state $\sigma_x$ on the other subsystem. We have
\begin{widetext} \begin{eqnarray}
P_x & = & \hbox{Tr}_{ab} \Big[
|\lambda\rangle\rangle\langle\langle \lambda|\: 1 \otimes
\Pi_x\Big] = (1-\lambda^2) \sum_q \: \lambda^{2q}\: \langle q|
\Pi_x | q\rangle = (1-\lambda^2)\hbox{Tr}_b\big[ \lambda^{2 b^\dag
b} \: \Pi_x \big] \label{pygen} \\ \varrho_x &
= & \frac{1}{P_x} \hbox{Tr}_b \Big[
|\lambda\rangle\rangle\langle\langle \lambda|\: 1 \otimes
\Pi_x\Big] = \frac{1-\lambda^2}{P_x} \sum_{pq} \: \lambda^{p+q} \:
\langle p | \Pi_x | q\rangle \: |p\rangle\langle q| = \frac{
\lambda^{a^\dag a} \:\Pi_x\: \lambda^{a^\dag a}}{\hbox{Tr}_b \big[
\lambda^{2 b^\dag b} \: \Pi_x \big]}\label{rhygen}
\end{eqnarray} \end{widetext}
where $\Pi_x$ is the probability measure (POVM) of the
measurement. In the last equalities of both Eqs. (\ref{pygen})
and (\ref{rhygen}) we have already performed the trace over the
Hilbert space ${\cal H}_a$. Also notice that in the last expression
of $\varrho_x$ $\Pi_x$ should be meant as an operator acting on
${\cal H}_a$. Our scheme is general enough to include the
possibility of performing any unitary operation on the beam
subjected to the measurement. In fact, if $E_x$ is the original
POVM and $V$ the unitary, the overall measurement process is
described by $\Pi_x=V^\dag E_x V$, which is again a POVM. In the following we
always consider $V=I$, {\em i.e.} no transformation before the measurement.
A further generalization consists in sending the result of the
measurement (by classical communication) to the reduced state
location and then performing a conditional unitary operation $U_x$
on the conditional state, eventually leading to the state
$\sigma_x = U_x \varrho_x U_x^\dag$. This degree of freedom will
be used in Section \ref{ss:heter}, where we analyze CV quantum
teleportation as a conditional measurement of the sum- and
difference-quadrature of two modes.
\section{Conditional measurements on twin-beam}
\label{s:details}
\subsection{Geigerlike ({\sc on/off}) photodetection}
\label{ss:onoff}
By looking at the expression (\ref{twbdef}) of TWB in the Fock
basis, it is apparent that ideal photocounting on one of the two
beams, described by the POVM $\Pi_n=|n\rangle\langle n|$, is a
conditional source of Fock number state $|n\rangle$, which would
be produced with a conditional probability $P_n=(1-\lambda^2)
\lambda^n$. However, as mentioned above, photocounting cannot be
considered a realistic kind of measurement.  Therefore, we now
consider the situation in which one of the two beams, say mode
$b$, is revealed by an avalanche {\sc on/off} photodetector, i.e.
a detector which has no output when no photon is detected and a
fixed output when one or more photons are detected. The action of
an {\sc on/off} detector is described by the two-value POVM
\begin{eqnarray}
\Pi_0\doteq\sum_{k=0}^\infty(1-\eta)^k|k\rangle\langle
k| \qquad\quad \Pi_1\doteq {{\bf I}} -\Pi_0\label{pom}\;
\end{eqnarray}
$\eta$ being the quantum efficiency. The outcome "1"
(i.e registering a "click" corresponding to one or more incoming
photons) occur with probability
\begin{eqnarray}
P_1 &=& \langle\langle\lambda | {{\bf I}} \otimes  \Pi_1 | \lambda\rangle\rangle
\\ &=& \frac{\eta \lambda^2}{1-\lambda^2 (1-\eta)}
 = \frac{\eta N}{2+\eta N}\nonumber
\label{probs}\;
\end{eqnarray}
and correspondingly, the conditional output states for
the mode $a$ is given by
\begin{eqnarray}
\varrho_1 = \frac{1-\lambda^2}{P_1}
\sum_{k=1}^\infty \lambda^{2k} \left[1-(1-\eta)^k\right]\: |k
\rangle\langle k|\label{fock}\;.
\end{eqnarray}
The density matrix in Eq. (\ref{fock}) describes a mixture: a {\em
pseudo}-thermal state where the vacuum component has been removed
by the conditional measurement. Such a state is highly
nonclassical, as also discussed in Ref. \cite{man}. Notice that
the nonclassicality is present only when the state exiting the
amplifier is entangled.  In the limit of low TWB energy the
conditional state $\varrho_1$ approaches the number state
$|1\rangle\langle 1|$ with one photon.  \par
The Wigner function
\begin{eqnarray}\label{Wrhodef}
W(\alpha)=\frac1{\pi^2} \int d^2\gamma \:
e^{\bar\gamma\alpha-\bar\alpha\gamma}\:\hbox{Tr}\left[\varrho_1\:
D(\gamma)\right]\:,
\end{eqnarray}
of $\varrho_1$ ($D(\gamma)=\exp[\gamma a^\dag -\bar\gamma a]$ is the displacement operator)
exhibits negative values for any value of $\lambda$ and $\eta$.
In particular, in the origin of the phase space we have
\begin{eqnarray}
W(0) =-\frac{2}{\pi}\:\frac{1}{N+1}\:\frac{2+\eta N}{2(1+N) -\eta N}
\label{wig0}\;.  \end{eqnarray}
One can see that also the generalized Wigner function for $s$-ordering
$$W_s(\alpha) = -\frac2{\pi s}\int d^2\: \gamma W_0(\gamma) \:
\exp\left[\:\frac2s\: |\alpha -\gamma|^2\right]\:,$$
shows negative values for $s \in (-1,0)$. In particular
one has
\begin{eqnarray}
W_s(0) = -
\frac{2(1+s)(2+\eta N)}{\pi(1+N-s)\left[2(1+N-s)-\eta N(1+s)\right]}
\label{ws0}\;.
\end{eqnarray}
A good measure of nonclassicality is given by the lowest index
$s^\star$ for which $W_s$ is a well-behaved probability, {\em
i.e.} regular and positive definite \cite{ncl}. Eq. (\ref{ws0})
says that for $\varrho_1$ we have $s^\star=-1$, that is
$\varrho_1$ describes a state as nonclassical as a Fock number
state.
\par The Fano factor of $\varrho_1$ is given by
\begin{eqnarray}
F=\frac12\: {\left( 2 + N \right) \, \left( 1 + {\frac{2}{2 +
N\,\eta}} - {\frac{4\,\left( 2 + N
\right) } {4 + N\,\left( 4 + N\,\eta
\right) }} \right)}
\label{fano}\;.
\end{eqnarray}
Roughly, for $\eta$ not too low, we have $F=N/2+ N(N-2)/(N+2)^2
(1-\eta)$. Therefore, we have that the beam $b$ is also
subPossonian for $N \lesssim 2.2$.  The verification of
nonclassicality can be performed, for any value of the gain, by
checking the negativity of the Wigner function through quantum
homodyne tomography \cite{robust}, and in the low gain regime,
also by verifying the subPoissonian character by measuring the
Fano factor via direct noise detection \cite{kum,garbo}.
\par
Notice that besides quantum efficiency, i.e. lost photons, the
performance of a realistic photodetector may be degraded by the
presence of dark-count, i.e. by "clicks" that do not correspond to
any incoming photon. In order to take into account both these
effects a real photodetector can be modeled as an ideal
photodetector (unit quantum efficiency, no dark-count) preceded by
a beam splitter (of transmissivity equal to the quantum
efficiency) whose second port is in an auxiliary excited state
({\em e.g.} a thermal state, or a random-phase coherent state),
which accounts for the background noise (thermal or Poissonian).
However, at optical frequencies the number of dark counts is
negligible and we are not going to take into account this effect,
which have been analyzed in details in Ref. \cite{robust}.
\par
We conclude that conditional {\sc on/off} photodetection on TWB provides
a reliable scheme to check nonclassical light. The nonclassicality, as well
as its verification, are robust against amplifier gain and detector efficiency.
\subsection{Homodyne detection}
\label{ss:homod}
In this Section we consider the kind of conditional state that can be obtained
by homodyne detection on one of the two beams exiting the NOPA. We will show
that they are squeezed states. We first consider ideal homodyne detection
described by the POVM $\Pi_x=|x\rangle\langle x|$ where
$$|x\rangle=\left(\frac2\pi \right)^{1/4} e^{-x^2} \sum_{n=0}^\infty
\frac{H_n(\sqrt{2} x)}{\sqrt{n!\: 2^n}} \: |n\rangle\:, $$ with  $H_n(x)$
denoting the $n$-th Hermite polynomials, is an eigenstate of the quadrature
operator $x_b=\frac12(b+b^\dag)$. Then, in the second part of the section we will
consider two kind of imperfections: non unit quantum efficiency and finite
resolution. As we will see, the main effect of the conditional measurement,
{\em i.e} the generation of squeezing, holds also for these realistic
situations.
\par
The probability of obtaining the outcome $x$ from a homodyne detection on
the mode $b$ is obtained from Eq. (\ref{pygen}). We have
\begin{eqnarray}
P_{x} = (1-\lambda)^2 \sum_{q=0}^\infty \lambda^{2q}\:
\left|\langle x | q \rangle \right|^2
=\frac{\exp\{-\frac{x^2}{2\sigma_{\lambda}^2}\}}
{\sqrt{2\pi\sigma_{\lambda}^2}}
\label{pyeta1}\;,
\end{eqnarray}
where
\begin{eqnarray}
\sigma_{\lambda}^{2}= \frac14 \frac{1+\lambda^{2}}{1-\lambda^{2}}
= \frac14  (1+N)\:.
\end{eqnarray}
$P_x$ is Gaussian with variance that increases as $\lambda$ is approaching unit.
In the (unphysical) limit $\lambda \rightarrow 1$ {\em i.e.} infinite gain of
the amplifier the distribution for $x$ is uniform over the real axis. The conditional
output state is given by Eq. (\ref{rhygen}), and since $\Pi_x$ is a pure POVM, it
is a pure state $\varrho_x=|\psi_x\rangle\langle \psi_x|$ where
\begin{eqnarray}
|\psi_x\rangle = \sqrt{\frac{1-\lambda^2}{P_x}}\:
\lambda^{a^\dag a} \: |x\rangle = \sum_k \psi_k \:| k \rangle
\label{rheta1}\;.
\end{eqnarray}
The coefficients of $|\psi_x\rangle$ in the Fock basis are given by
\begin{eqnarray}
\psi_k = \left(\frac{\lambda^2}{2}\right)^{k/2}\!\!\!\!
\frac{1}{\sqrt{k!}}\:(1-\lambda^4)^{\frac14}
e^{-\frac{2\lambda^2 x^2}{1+\lambda^2}}\:H_k (\sqrt{2}x)\:,
\end{eqnarray}
which means that $|\psi_x\rangle$ is a squeezed state of the form
\begin{eqnarray}
|\psi_x\rangle & = & D(\alpha) S(\zeta) |0\rangle \label{sqy}
\:,\end{eqnarray}
where
\begin{eqnarray}
\alpha  &=&   \frac{2x\lambda}{1+\lambda^2} =   \frac{x\sqrt{N(N+2)}}{1+N} \nonumber \\
\zeta &=& \hbox{ArcTanh} \lambda^2 = \hbox{ArcTanh}\: \frac{N}{N+2}\:.
\end{eqnarray}
The quadrature fluctuations are given by
\begin{eqnarray}
\overline{\Delta x_a^2} = \frac14 \frac1{1+N} \qquad
\overline{\Delta y_a^2} = \frac14 (1+N) \label{fluct}\:,
\end{eqnarray}
where $x_a=\frac12(a^\dag +a)$, $y_a=\frac{i}2(a^\dag -a)$, and
$\overline{\Delta O^2}=\hbox{Tr}[\varrho \: O^2] -
(\hbox{Tr}[\varrho\: O])^2$. Eq. (\ref{fluct}) confirms that
$|\psi_x\rangle$ is a minimum uncertainty state. Notice that: i)
the amount of squeezing is independent on the outcome of the
measurement, which only influences the coherent amplitude; ii)
according to Eq. (\ref{pyeta1}) the most probable conditional
state is a  squeezed vacuum. The average number of photon of the
conditional state is given by
\begin{eqnarray}
N_x=\langle \psi_x| a^\dag a | \psi_x\rangle = x^2 \frac{N(N+2)}{(1+N)^2}
+\frac14 \frac{N^2}{1+N}\:.
\end{eqnarray}
The conservation of energy may be explicitly checked by averaging over
the possible outcomes
\begin{eqnarray}
\int dx \: P_x \: N_x = \frac14 \frac{N^2}{1+N} + \sigma_\lambda^2
\: \frac{N(N+2)}{(1+N)^2} = \frac{N}{2} \label{checkE}\:,
\end{eqnarray}
which correctly reproduces the number of photon pertaining each part of the TWB.\par
We now take into account the effects of nonunit quantum efficiency at the
homodyne detector on the conditional state. We anticipate that $\varrho_x$ will
be no longer pure states, and in particular they will not be
squeezed states of the form (\ref{sqy}). Nevertheless, the conditional output
states still exhibit squeezing {\em i.e.} quadrature fluctuations below the coherent
level, for any value of the outcome $x$, and for quantum efficiency larger
than $\eta > 1/2$.
\par
\begin{figure}[h]
\begin{tabular}{lll}
\psfig{file=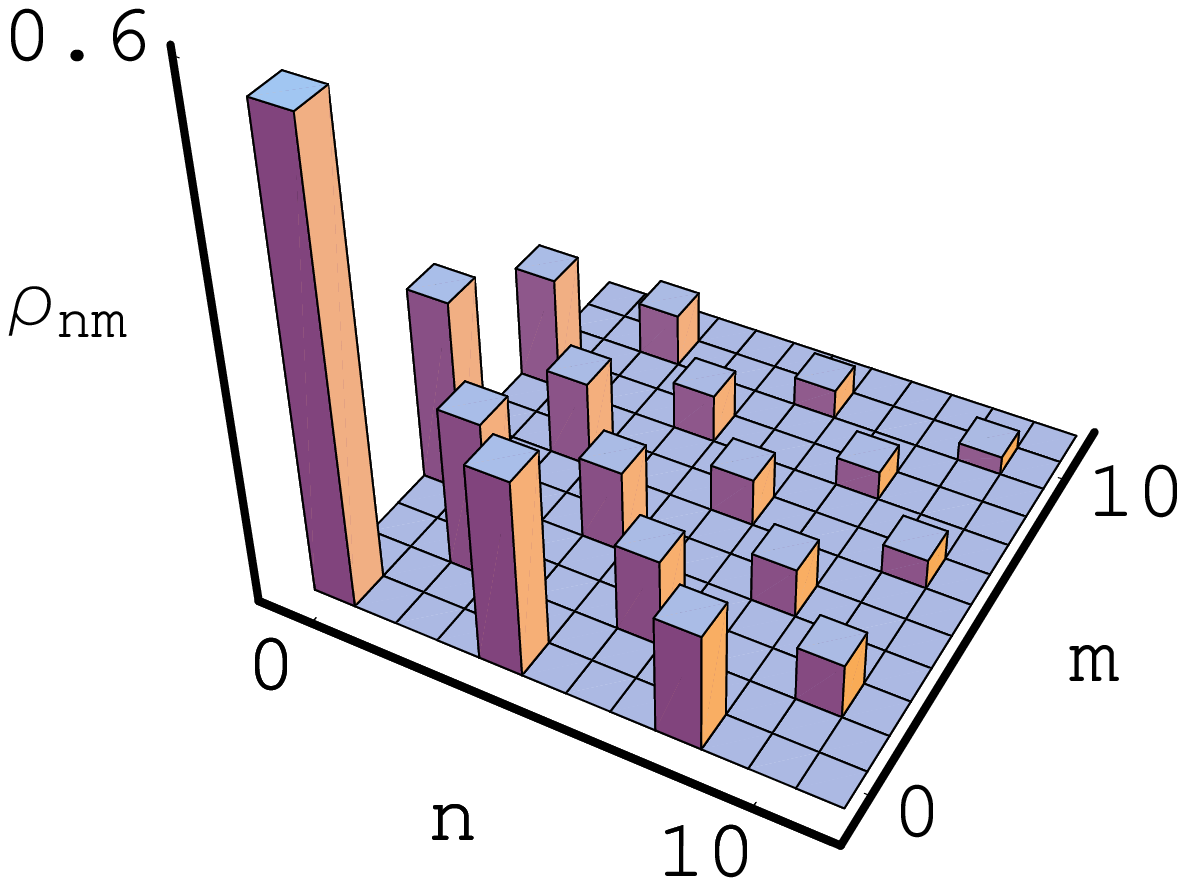,width=5cm} &
\psfig{file=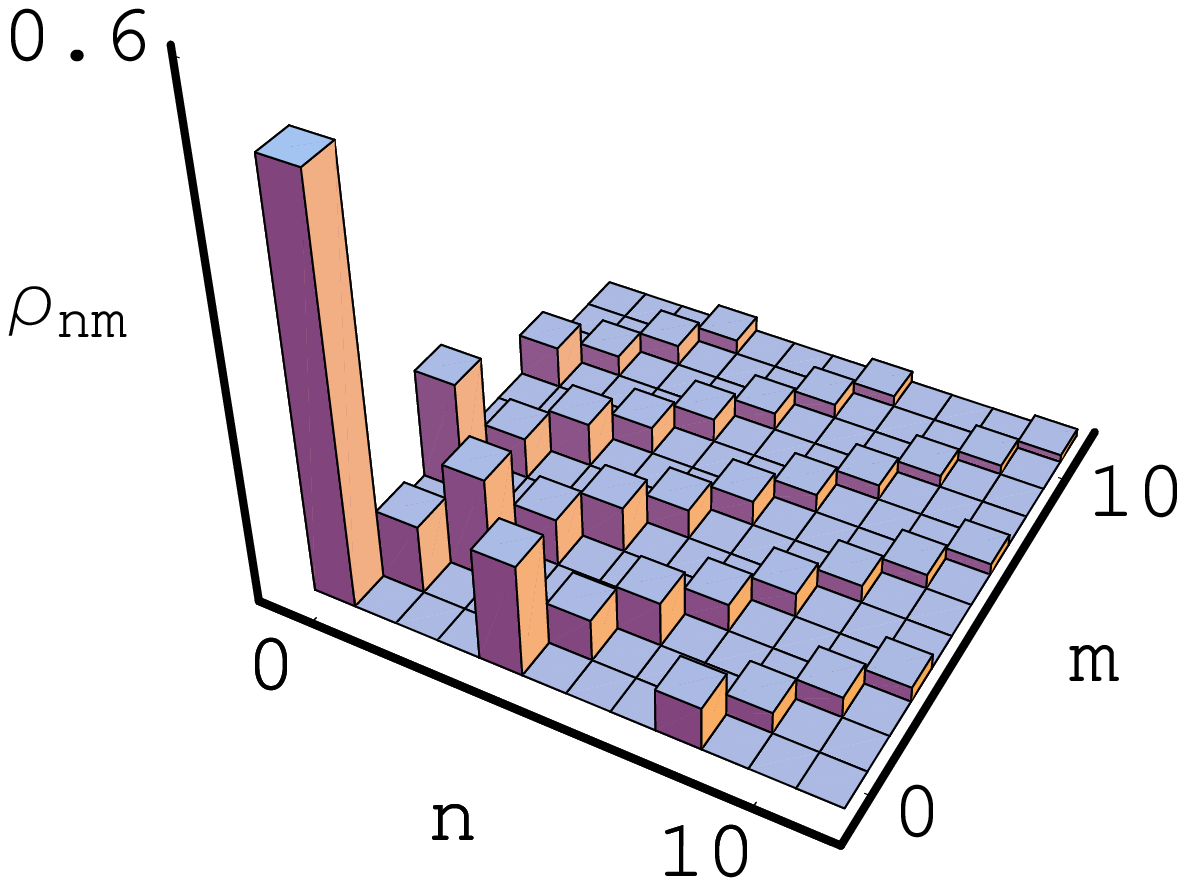,width=5cm} &
\psfig{file=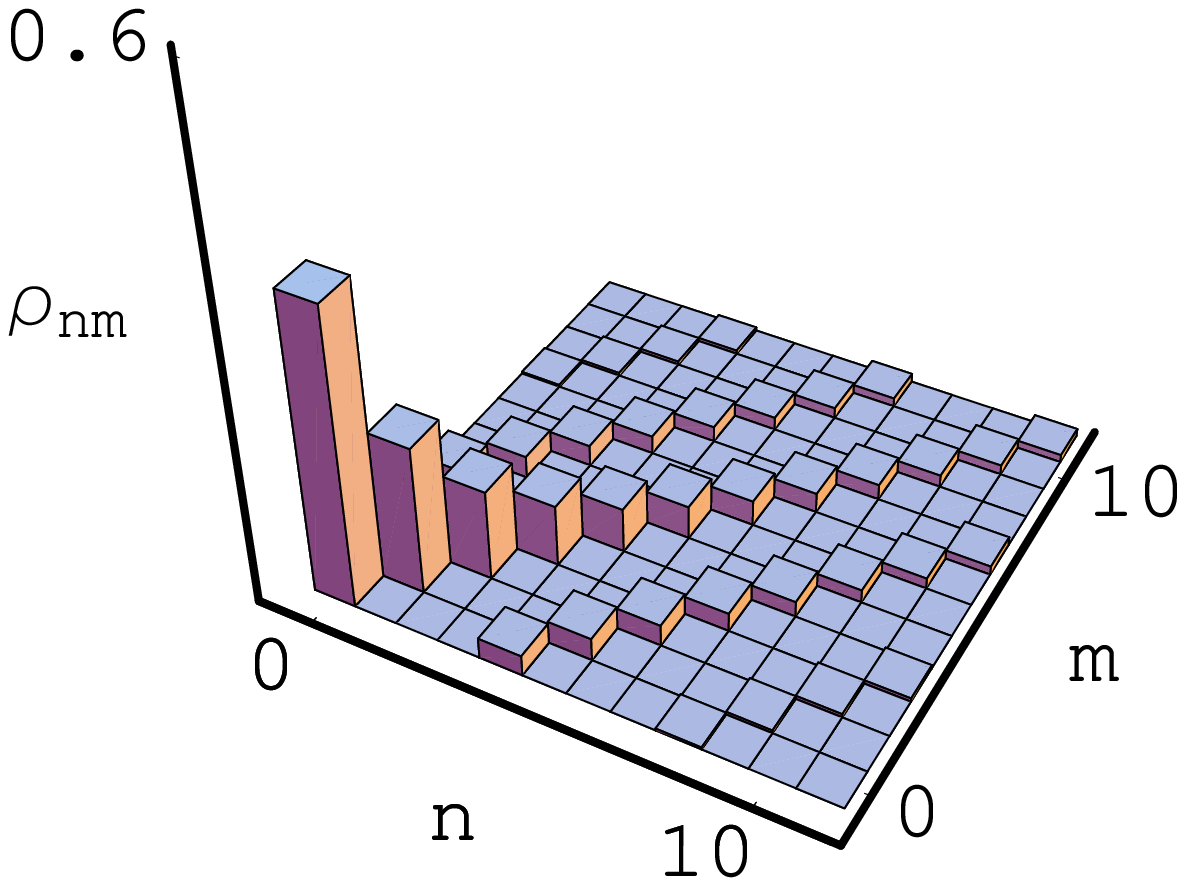,width=5cm} \\
\psfig{file=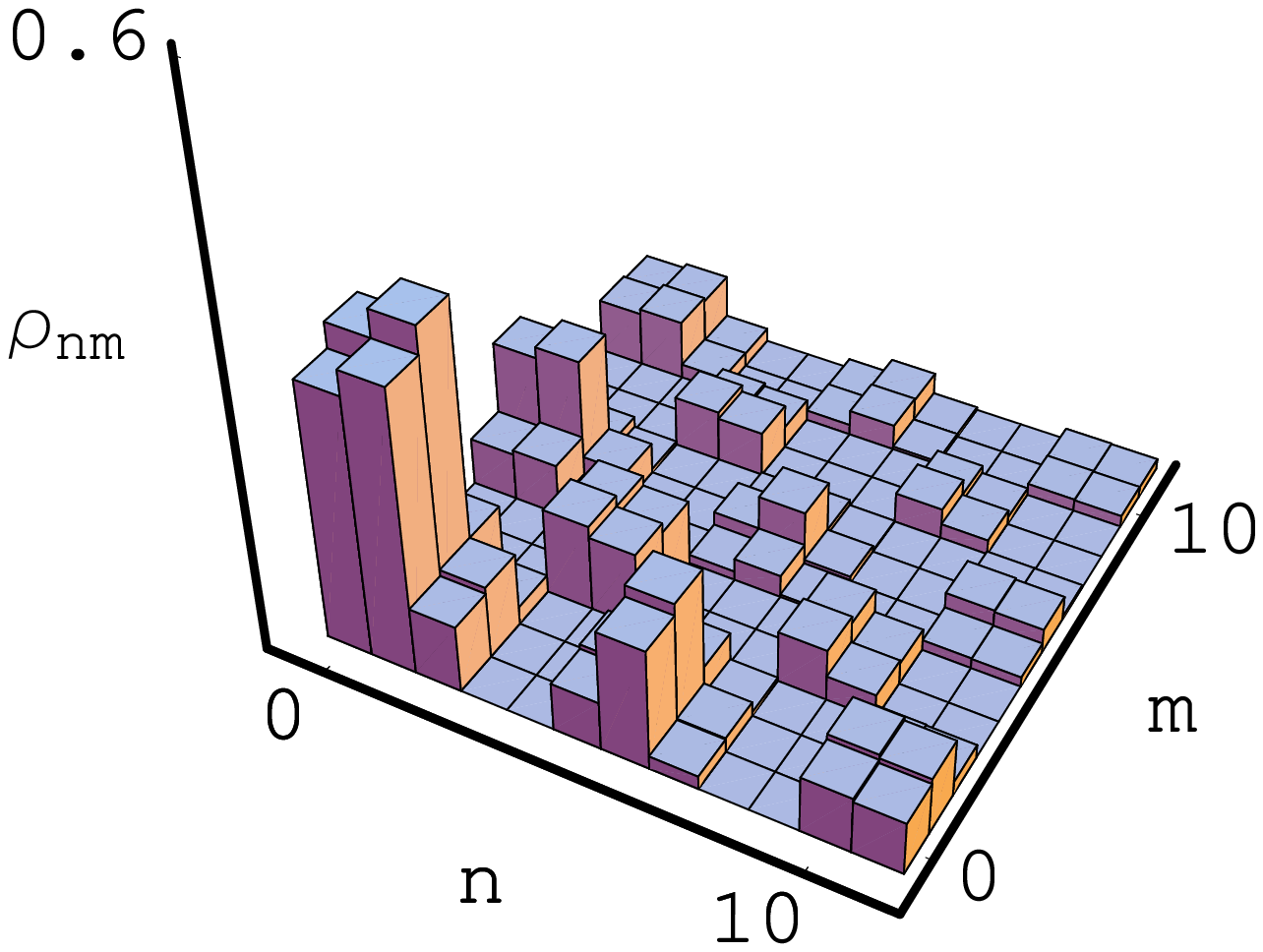,width=5cm} &
\psfig{file=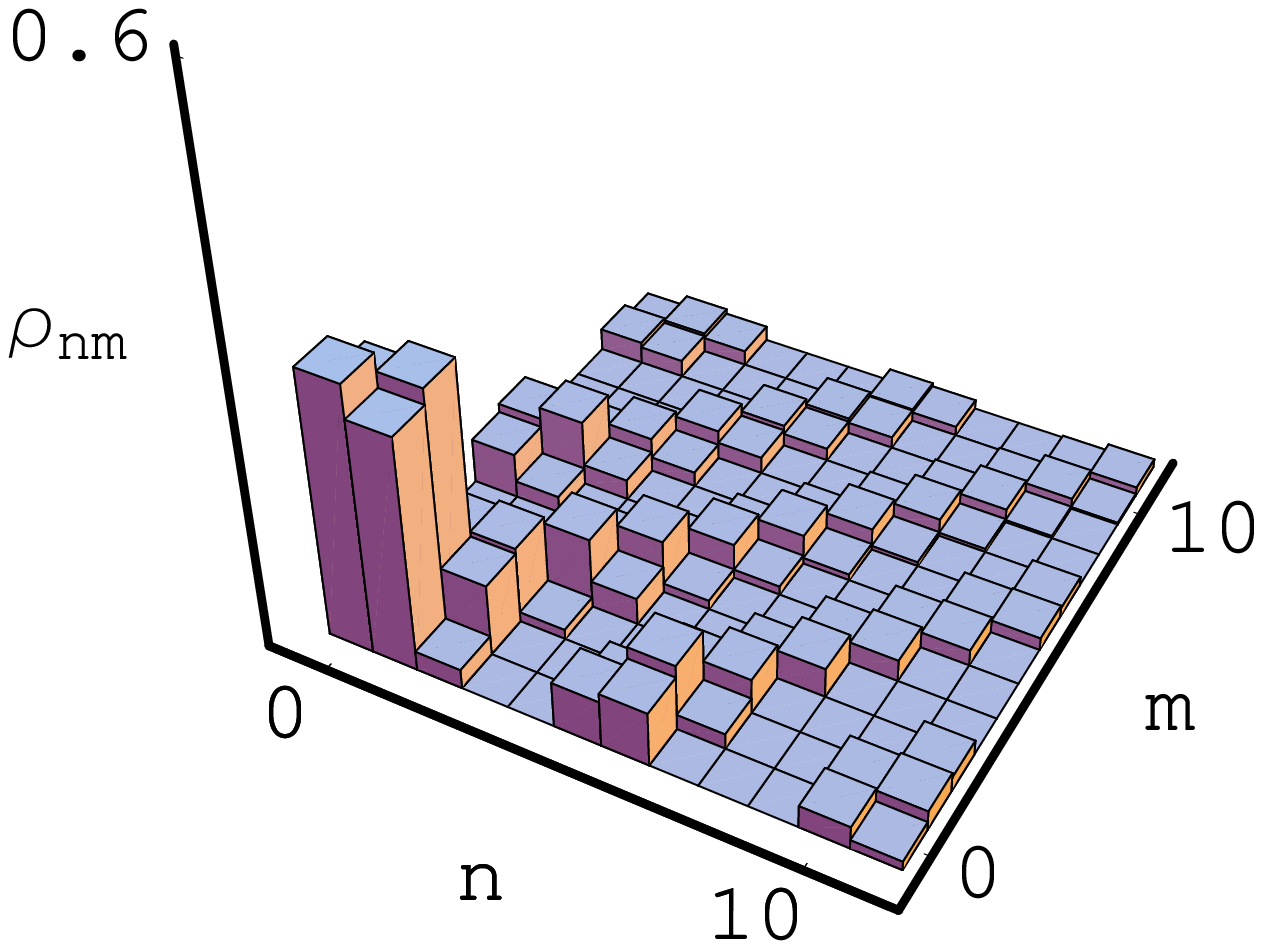,width=5cm} &
\psfig{file=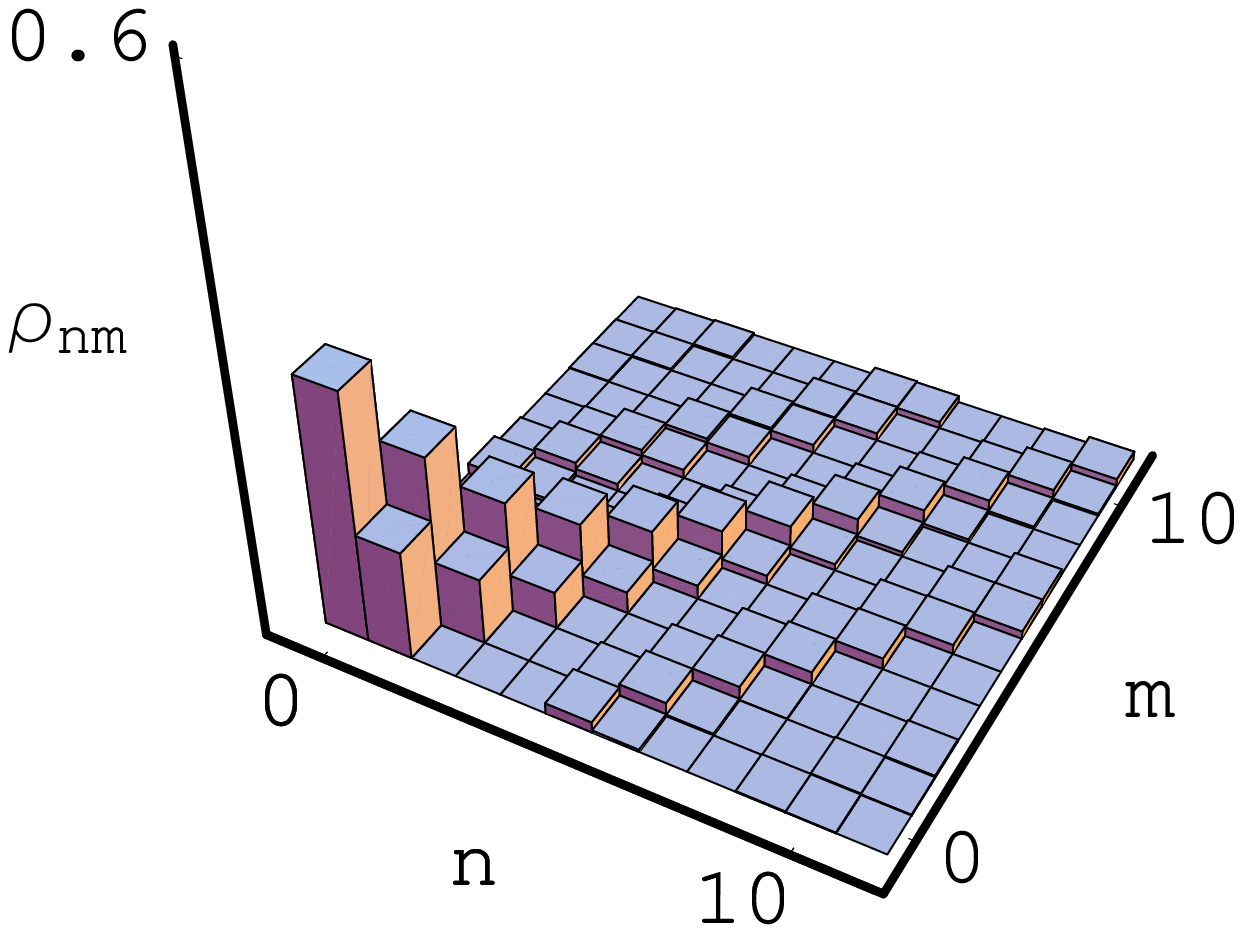,width=5cm}
\end{tabular}
\caption{Matrix elements in the Fock basis of the conditional state
$\varrho_x$ after homodyne detection on TWB. In the first row the matrix
elements for $x=0.0$ and $\eta=1.0,0.8,0.4$. In the second row the matrix
elements for $x=0.6$ and the same values of quantum efficiency.}
\label{f:pdn}
\end{figure}

The POVM of a homodyne detector with quantum efficiency $\eta$ is given by
\begin{eqnarray}
\Pi_{x\eta}=\int
\frac{dt}{\sqrt{2\pi\Delta_{\eta}^{2}}}\exp\left\{-
\frac{(x-t)^{2}}{2\Delta_{\eta}^{2}}\right\}\:\Pi_t
\label{Piyeta}\;,
\end{eqnarray}
where
\begin{eqnarray}
\Delta_{\eta}^{2} = \frac{1-\eta}{4 \eta}\:.
\label{Deltaeta}
\end{eqnarray}
The nonideal POVM is a Gaussian convolution of the ideal POVM. The
main effect is that $\Pi_{x\eta}$ is no longer a pure orthogonal
POVM. The probability $P_{x\eta}$ of obtaining the outcome $x$ is
still a Gaussian, now with  variance
\begin{equation}
\Delta^{2}_{\lambda\eta} = \sigma_{\lambda}^{2} + \Delta_{\eta}^{2}\:.
\label{totdl}
\end{equation}
The conditional output state is again given by Eq. (\ref{rhygen}).
After some algebra we get the matrix element in the Fock basis
\begin{widetext} \begin{equation}
\langle n | \varrho_x | m \rangle = \frac{\left(1-\lambda^2\right)
\lambda^{n+m}}{\sqrt{n!m!2^{n+m}}}
\sqrt{\eta\:\frac{2-\eta(1-\lambda^2)}{1-\lambda^2}}\:e^{-4
x^2\frac{\eta^2\lambda^2}{1-\lambda^2(1-2\eta)}}\:
\sum_{k=0}^{min(m,n)}2^{k}k! {m \choose k} {n \choose
k}\eta ^{\frac{m+n}2-k}
H_{m+n-2k}\left(\sqrt{2\eta}\:x \right)
\:.\end{equation}
\end{widetext}
The quadrature fluctuations are now given by
\begin{eqnarray}
\overline{\Delta x_a^2} = \frac{1+N(1-\eta)}{4(1+\eta N)} \qquad
\overline{\Delta y_a^2} = \frac14 (1+N) \label{fleta}
\:.\end{eqnarray} As a matter of fact, $\overline{\Delta y_a^2}$
is independent on $\eta$, whereas $\overline{\Delta x_a^2}$
increases for decreasing $\eta$. Therefore, the conditional output
$\varrho_x$ is no longer a minimum uncertainty state. However, for
$\eta$ large enough we still observe squeezing in the direction
individuated by the measured quadrature. The form of the output
state can be obtained by the explicit calculation of the matrix
elements or, more conveniently, by evaluating the Wigner function 
(see Section \ref{s:wig}). We have
\begin{eqnarray}
\varrho_{x\eta}= D(\alpha_{\eta})\:S(\zeta_{\eta})\: \nu_{th}\: S^\dag(\zeta_{\eta})
\:D^\dag (\alpha_{\eta})
\label{outxeta}\;,
\end{eqnarray}
where $$\nu_{th}= (1+n_{th})^{-1} \sum_{p=0}^\infty
\left(\frac{n_{th}}{1+n_{th}}\right)^p \:
|p\rangle\langle p|$$ is a thermal state with average number of photons given
by
\begin{eqnarray}
n_{th} &=& \frac12 \left\{\sqrt{\frac{(1+N)[1+N(1-\eta)]}{1+\eta N}}-1\right\}
\label{ntheta}\;,
\end{eqnarray}
and the amplitude and squeezing parameters read as follows
\begin{eqnarray}
\alpha_{\eta} &=& \frac{\eta\sqrt{N(N+2)}}{1+\eta N}\: x
\\  \zeta_{\eta} &=& \frac14 \log \frac{(1+N)(1+\eta N)}{1+N(1-\eta)}
\label{rxeta}\;.
\end{eqnarray}
From Eqs. (\ref{fleta}) and (\ref{rxeta}) we notice that $\varrho_{x\eta}$
shows squeezing if $\eta > 1/2$, independently on the actual value $x$
of the homodyne outcome.
In Fig. \ref{f:pdn} we illustrate the effects of quantum efficiency on the
matrix elements of the conditional state. In particular we plot the matrix
elements for two values of the homodyne outcome $x=0.0,0.6$ and three values
of the quantum efficiency $\eta=1.0,0.8,0.4$.
\par
\begin{figure}[h]
\begin{turn}{-90}
\psfig{file=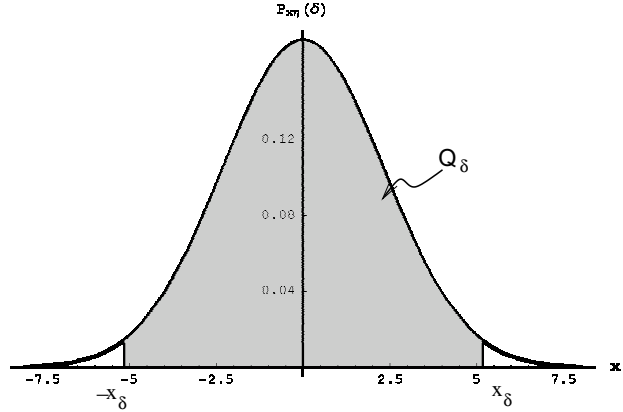,width=55mm}
\end{turn}
\caption{Probability distribution $P_{x\eta}(\delta)$ of the homodyne
outcomes $x$ for $\eta=0.7$, $N=20$, and $\delta=0.25$. The threshold
value $x_\delta\simeq 5.16$ to obtain a conditionally squeezed state
is shown. The gray-shaded area represents the overall probability
$Q_\delta\simeq 97 \%$ of producing a squeezed state by the conditional
measurement.}\label{f:probxd}
\end{figure}
\begin{figure}[h]
\psfig{file=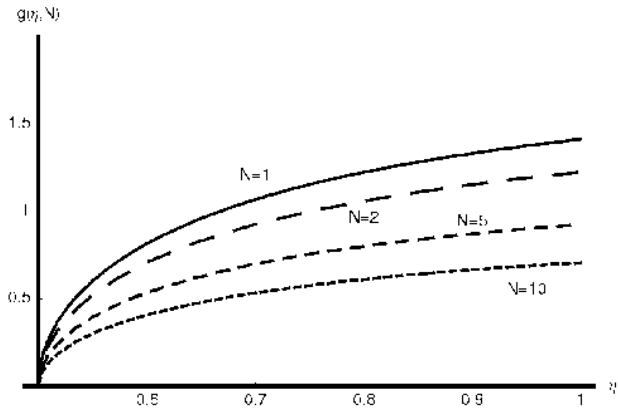,width=90mm}
\caption{The function $g(\eta,N)$ in Eq.(\ref{getaN}) versus the
quantum efficiency for different values of the TWB photon number
$N$. From top to bottom we have the curves for
$N=1,2,5,10$.}\label{f:psq}
\end{figure}

The outcome of homodyne detection are, in principle, continuously
distributed over the real axis. However, in practice, one has
always to discretize data, mostly because of finite experimental
resolution. The POVM describing homodyne detection with binned
data is given by
\begin{eqnarray}
\Pi_{x\eta}(\delta) = \frac1\delta \int_{x-\delta/2}^{x+\delta/2} dt \:
\Pi_{t\eta} \label{binnedPOVM}\;,
\end{eqnarray}
where $\Pi_{t\eta}$ is given in Eq.(\ref{Piyeta}),
and $\delta$ is the width of the bins.
The probability distribution is now given by
\begin{eqnarray}
P_{x\eta}(\delta) &=& \frac{1}{2\delta}
\left[ \hbox{Erf}\left(\frac{x+\frac\delta2}{\sqrt{2\Delta^2_{\lambda\eta}}}\right)
-\hbox{Erf}\left(\frac{x-\frac\delta2}{\sqrt{2\Delta^2_{\lambda\eta}}}\right)
\right] \\ &=&\frac{\exp\left\{ -\frac{x^2}{2\Delta^2_{\lambda\eta}}\right\}}
{\sqrt{2\pi\Delta^2_{\lambda\eta}}}\:
\left\{1-\frac{x^2-\Delta^2_{\lambda\eta}}{24\Delta^2_{\lambda\eta}}
\:\delta^2 \right\} + O(\delta^3) \nonumber
\label{probrefin}\;
\end{eqnarray}
where $\Delta_{\lambda\eta}^2$ is given in Eq. (\ref{totdl}) and
$\hbox{Erf}[...]$ denotes the error function. The conditional
state is modified accordingly. Concerning the quadrature
fluctuations of the conditional state we have, up to second order
in $\delta$,
\begin{eqnarray}
\overline{\Delta x_a^2}(\delta) = \overline{\Delta x_a^2} +
x^2 \frac{\delta^2}{12} \frac{\eta^2 N (N+2)}{(1+\eta N)^2}
\label{flbin}\;,
\end{eqnarray}
which is below the coherent level for $\eta > 1/2$ and for
\begin{eqnarray}
|x| < x_\delta \equiv
\frac1\delta \sqrt{\frac{3(1+\eta N)(2\eta-1)}{\eta^2(N+2)}}
\label{xdel}\;.
\end{eqnarray}
Therefore, the effect of finite resolution is that the conditional
output is squeezed only for the subset $|x|< x_\delta$ of the possible
outcomes which, however, represents the range where the probability is higher.
In Fig. \ref{f:probxd}, as an example, we show $P_{x\eta}(\delta)$ as a function of $x$
for $\eta=0.7$, $\delta=0.25$, and $N=20$. The threshold $x_\delta$ is
shown as well as the overall probability $Q_\delta$ of producing a squeezed
state which, up to second order in $\delta$, is given by
\begin{eqnarray}
Q_\delta= \int_{-x_\delta}^{x_\delta} dx\: P_{x\eta} (\delta) =
\left\{\begin{array}{ll}
0 & N=0 \\
\hbox{Erf}
\left[\:\frac1\delta \: g(\eta,N)\right] & N\neq 0
\end{array}\right.
\label{probxd}\;,
\end{eqnarray}
where
\begin{eqnarray}
g(\eta,N)= \sqrt{\frac{6 (2\eta-1)}{\eta(N+2)}}
\label{getaN}\;.
\end{eqnarray}
In Fig. \ref{f:psq} we show $g(\eta,N)$ as a function of $\eta$
for different values of the TWB photon number $N$. As it is
apparent from the plot $g(\eta,N)$ is a monotonically increasing
function of $\eta$ and a monotonically decreasing function of
$N$. Notice that the larger is $g(\eta,N)$ the smaller is the
effect of finite resolution in decreasing the probability of
obtaining squeezed states. In principle, using small value of $N$
({\em i.e} less entanglement) increases the probability of
getting squeezed states. However, such states would be only
slightly squeezed {\em i.e.} $\overline{\Delta x^2_a}\lesssim
\frac14$. Therefore, since the scheme is aimed to be a tunable
source of  squeezing, the best strategy is to use large values of
$N$, while accepting a slightly decreased conditional
probability.
\subsection{Joint measurement of two-mode quadratures}
\label{ss:heter} In this Section we assume that mode $b$ is
subjected to the measurement of the the real and the imaginary
part of the complex operator $Z=b+c^\dag$, where $c$ is an
additional mode excited in a reference state $S$. The measurement
of $\hbox{Re}[Z]$ and $\hbox{Im[Z]}$ corresponds to measuring the
sum- and difference-quadratures $x_b+x_c$ and $y_b-y_c$ of the two
modes, and can be experimentally implemented by multiport homodyne
detection, if the two modes have the same frequencies
\cite{msa,rip,tri}, or by heterodyne detection otherwise
\cite{het}. The measurement is described by the following POVM
\cite{bpl,ics7}
\begin{equation}\label{POVM:tele}
{\Pi}_\alpha = \frac{1}{\pi}
{D}(\alpha) \, {S}^{T} \, {D}^{\dag}(\alpha)
\end{equation}
where $\alpha$ is a complex number, $D(\alpha)$ is the
displacement operator and $(\cdots)^{T}$ stands for the
transposition operation.  The present scheme is equivalent to that
of CV teleportation, which can be viewed as a conditional
measurement, with the state to be teleported playing the role of
the reference state $S$ of the apparatus. In order to complete the
analogy we assume that the result of the measurement is
classically transmitted to the receiver's location, and that a
displacement operation $D(\alpha)$ is performed on the conditional
state $\varrho_\alpha$.  One has
\begin{eqnarray}
p_\alpha &=& (1-\lambda^2) \hbox{Tr}_1 \left[
\lambda^{a^\dag a} \: D(\alpha) S^T D^\dag (\alpha)
\right]\\
\varrho_\alpha &=& \frac1{p_\alpha} \hbox{Tr}_1
\left[|\lambda\rangle\rangle\langle\langle \lambda|\:
D(\alpha) S^T D^\dag (\alpha)
\otimes I_2\right] \nonumber \\
\sigma_\alpha &=& D(\alpha)\varrho_\alpha D^\dag (\alpha)\nonumber
\label{condalpha}\;,
\end{eqnarray}
while the teleported state is the average over all 
the possible outcomes, {\em i.e.}
\begin{widetext}
\begin{eqnarray}
\sigma = \int d^2\alpha \: p_\alpha \: \sigma_\alpha
= \int d^2\alpha \:
D(\alpha)\:\hbox{Tr}_1
\left[|\lambda\rangle\rangle\langle\langle \lambda|\:
D(\alpha) S^T D^\dag (\alpha)
\otimes I_2\right] D^\dag (\alpha)
\label{teleported}\;.
\end{eqnarray}
\end{widetext}
After performing the partial trace, and some algebra, one has
\begin{eqnarray}
\sigma = \int \frac{d^2\alpha}{\pi K_0}\: \exp\{-\frac{|\alpha|^2}{K_0}\}
\: D(\alpha)S D^\dag (\alpha)
\label{tel}\;,
\end{eqnarray}
where $K_0=1+N-\sqrt{N(N+2)}$. The output state $\varrho$
coincides with the input only in the limit $N\longrightarrow
\infty$ {\em i.e.} for infinite energy of the TWB. Eq. (\ref{tel})
shows that CV teleportation with finite amount of entanglement is
equivalent to a Gaussian channel with $K_0$ background photons
applied to the input state. This result has been also obtained in Refs. 
\cite{hof,ban} by different methods. In the next Section we
will show that this result still holds taking into the effects of 
decoherence due to losses, and nonunit quantum efficiency of the measurement,
either multiport homodyne or heterodyne detection.
\section{Conditional measurements in the phase space}
\label{s:wig}
The results of the previous Sections can be derived, and for CV
teleportation also extended, using Wigner functions in the phase
space. The analysis is based on the fact that the trace
between two operators can be written as \cite{cah}
\begin{eqnarray}
\hbox{Tr}\left[O_1\: O_2\right] &=& \pi \int
d^2\beta\:W[O_1](\beta)\:W[O_2](\beta) \label{trax}\;,
\end{eqnarray}
where the Wigner function for a generic operator $O$ is defined
analogously to that of a density matrix. As in (\ref{Wrhodef}) we
write
\begin{eqnarray}
W[O](\alpha) = \int \frac{d^2\gamma}{\pi^2}\: e^{\alpha\bar\gamma
- \bar\alpha\gamma}\: \hbox{Tr}\left[O\: D(\gamma)\right]
\label{Wigs}\;,
\end{eqnarray}
where $\alpha$ is a complex number and $D(\gamma)$ is the
displacement operator. The inverse transformation reads as follows
\cite{msacchi}
\begin{eqnarray}
O =  \int d^2\alpha \: W[O] (\alpha)\:  e^{-2 |\alpha|^2} \:
e^{2\alpha a^\dag} \left(-\right)^{a^\dag a} e^{2\bar\alpha a}
\label{invwig}\;.
\end{eqnarray}
The Wigner function $W[\lambda](x_1,y_1;x_2,y_2)$ of a TWB is
Gaussian (we omit the argument)
\begin{widetext}
\begin{eqnarray}
W[\lambda]=\left(2\pi \sigma_+^2 \: 2\pi
\sigma_-^2\right)^{-1}\:
\exp\left[-\frac{(x_1+x_2)^2}{4\sigma_+^2}
-\frac{(y_1+y_2)^2}{4\sigma_-^2} -\frac{(x_1-x_2)^2}{4\sigma_-^2}
-\frac{(y_1-y_2)^2}{4\sigma_+^2}\right]
\label{wtwb}\;
\end{eqnarray}
\end{widetext}
where the variances are given by
\begin{eqnarray}
\sigma^2_+ &=&\frac14\left[1+N+\sqrt{N(N+2)}\: \right] \\
\sigma^2_- &=&\frac14\left[1+N-\sqrt{N(N+2)}\: \right]
\label{stwb}\;.
\end{eqnarray}
Using (\ref{trax}) we rewrite the probability distribution
(\ref{pygen}) as follows
\begin{widetext}
\begin{eqnarray}
P_x &=& \!\!\int\!\!\!\!\int\!\!  dx_1 dy_1
 \int\!\!\!\!\int\!\! dx_2 dy_2 \:
W[\lambda](x_1,y_1;x_2,y_2)\: W[\Pi_x] (x_2,y_2) \\
&=& (1-\lambda^2)
 \int\!\!\!\!\int\!\! dx_2 dy_2 \:
W[\lambda^{2b^\dag b}](x_2,y_2)\: W[\Pi_x] (x_2,y_2)
\label{pgenW}\;,
\end{eqnarray}
\end{widetext}
where $W[\Pi_x](x_2,y_2)$ is the Wigner function of the POVM describing the
measurement and
\begin{eqnarray}
W[\lambda^{2b^\dag b}](x_2,y_2) = \frac1{\pi(1+N)}\exp
\left(-\frac{x_2^2+y_2^2}{1+N}\right)\label{w2l}\;.
\end{eqnarray}
Analogously, the Wigner function of the conditional output state (\ref{rhygen})
can be written as
\begin{widetext}
\begin{eqnarray}
W[\varrho_x](x_1,y_1) = \frac{1}{P_x}\: \int\!\!\!\!\int\!\! dx_2 dy_2 \:
W[\lambda](x_1,y_1;x_2,y_2)\: W[\Pi_x] (x_2,y_2)
\label{rhgenW}\;.
\end{eqnarray}
\end{widetext}
Once the Wigner function for the POVM $\Pi_x$ of the detector is
known, one may reproduce the results of previous Sections using
(\ref{pgenW}) and (\ref{rhgenW}) together with (\ref{invwig}). For
{\sc on/off} photodetection one has
\begin{eqnarray}
W[\Pi_0](x_2,y_2) &=& \frac{2}{\pi(2-\eta)}
\exp\left(-2\:\frac{x_2^2+y_2^2}{2-\eta}\right)
\nonumber \\
W[\Pi_1](x_2,y_2) &=& 1- W[\Pi_0](x_2,y_2)
\label{onoffW}\;,
\end{eqnarray}
whereas the POVM of a homodyne detector with quantum
efficiency $\eta$ corresponds to the Wigner function given by
\begin{eqnarray}
W[\Pi_{x\eta}] (x_1,y_1) &\equiv& W[\Pi_{x\eta}] (x_1)
\nonumber \\ &=& (2\pi\Delta_\eta^2)^{-1/2}
\exp\left\{-\frac{(x_1-x)^2}{2\Delta_\eta^2} \right\}
\label{wighometa}\;,
\end{eqnarray}
where $\Delta_\eta^2$ is given in Eq. (\ref{Deltaeta}). \par Let
us now focus our attention on the situation where the conditional
measurement on TWB is the joint measurement of the sum- and
difference-quadratures of two modes. In this case, the Wigner
approach is advantageous, in particular in the description of
optical teleportation as a conditional measurement, in order to
include the degrading effects of nonunit quantum efficiency and of
losses along the transmission channel. \par 
At first, we consider the ideal POVM $\Pi_\alpha$ of Eq. 
(\ref{POVM:tele}). By taking into
account that for any density matrix
\begin{eqnarray}
W[\varrho^T](x,y) &=& W[\varrho](x,-y) \nonumber \\
W[D(\alpha)\varrho D^\dag (\alpha)](x,y) &=&
W[\varrho](x-x_\alpha,y-y_\alpha) \label{rems}\;,
\end{eqnarray}
with $x_\alpha=\hbox{Re} [\alpha]$ and
$y_\alpha=\hbox{Im}[\alpha]$, it is easy to show that
\begin{eqnarray}
W[\Pi_\alpha](x_2,y_2) = W[S](x_2-x_\alpha,y_\alpha-y_2)
\label{Wpomalfa}\;.
\end{eqnarray}
Inserting (\ref{Wpomalfa}) in (\ref{pgenW}) and (\ref{rhgenW}),
and changing the integration variables, we obtain the Wigner function
of the teleported state $\sigma$ of Eq. (\ref{condalpha})
\begin{widetext}
\begin{eqnarray}
W[\sigma](x_2,y_2) &=&
\!\!\int\!\!\!\!\int\!\!  dx_1 dy_1
 \int\!\!\!\!\int\!\! dx_\alpha dy_\alpha
\:  W[\lambda](x_2+x_\alpha,y_2+y_\alpha;x_1+x_\alpha,-y_1-y_\alpha)
\: W[S] (x_1,y_1) \nonumber \\
&=& \!\!\int\!\!\!\!\int\!\!  \frac{dx_1 dy_1}{\pi K_0}
\exp\left\{-\frac{x_1^2+y_1^2}{K_0}\right\}\: W[S]
(x_2-x_1,y_2-y_1)\label{wigtel}\;,
\end{eqnarray}
\end{widetext}
which corresponds to the state given by Eq. (\ref{tel}).
We now proceed by taking into account nonunit quantum efficiency
of the detector and losses due to propagation of TWB.
Nonunit quantum efficiency at either double homodyne or heterodyne
detectors modifies the POVM of the sender, which becomes a Gaussian
convolution of the ideal POVM $\Pi_\alpha$
\begin{eqnarray}
\Pi_{\alpha\eta} = \int \frac{d^2\beta}{\pi D_\eta^2} \:
\exp\{-\frac{|\alpha - \beta|^2}{D_\eta^2}\} \: \Pi_\beta
\qquad D^2_\eta=\frac{1-\eta}{\eta} \label{pialfaeta}\;,
\end{eqnarray}
leading to
\begin{widetext}
\begin{eqnarray}
W[\Pi_{\alpha\eta}](x_2,y_2) =
\!\int\!\!\!\int\!  \frac{dx_\beta dy_\beta}{\pi D_\eta^2}
\exp\left(-\frac{x_\beta^2+y_\beta^2}{D_\eta^2}\right)
W[S](x_\beta+x_2-x_\alpha,y_\alpha-y_2-y_\beta)
\label{Wpomalfaeta}\;.
\end{eqnarray}
On the other hand, losses that may occur during the propagation of TWB
degrade the entanglement. This effect can be described as the coupling
of each part of the TWB with a non zero temperature reservoir.
The dynamics is described by the two-mode Master equation
\begin{eqnarray}  \frac{d R_t}{dt} \equiv {\cal L}
R_t = \Gamma (1+M) L[a] R_t + \Gamma (1+M) L[b] R_t
+ \Gamma M L[a^\dag ] R_t + \Gamma  M  L[b^\dag] R_t
\label{master} \end{eqnarray}
\end{widetext}
where $R_t\equiv R(t)$, $R_0=|x\rangle\rangle\langle\langle x|$,
$\Gamma$ denotes the (equal) damping rate, $M$ the number of
background thermal photons, and $L[O]$ is the Lindblad
superoperator $L[ O ] \sigma_t =  O \sigma_t O^\dag - \frac{1}{2}
O^\dag  O \sigma_t - \frac{1}{2} \sigma_t OO^\dag\:.$ The terms
proportional to $L[a]$ and  $L[b]$ describe the losses, whereas
the terms proportional to $L[a^\dag]$ and $L[b^\dag]$ describe a
linear phase-insensitive amplification process. This can be due
either to optical media dynamics or to thermal hopping; in both
cases no phase information is carried. Of course, the dissipative
dynamics of the two channels are independent on each other. The
Master equation (\ref{master}) can be transformed into a
Fokker-Planck equation for the two-mode Wigner function of the
TWB. Using the differential representation of the superoperators
in Eq. (\ref{master}) the corresponding Fokker-Planck equation
reads as follows
\begin{widetext}
\begin{eqnarray}
\partial_\tau W_\tau = \left[ \frac{1}{8}
\left(\sum_{j=1}^2\partial^2_{x_j x_j} + \partial^2_{y_j
y_j}\right) + \frac{\gamma}2 \left(\sum_{j=1}^2 \partial_{x_j}
x_j+\partial_{y_j}  y_j + \right) \right] W_\tau \label{fp}\:,
\end{eqnarray}
where $\tau$ denotes the rescaled time $\tau=\Gamma/\gamma\:t$,
and $\gamma= \frac{1}{2M+1}$ the drift term. The solution of Eq.
(\ref{fp}) can be written as
\begin{eqnarray}
W_\tau &=& \int _{}dx^{\prime}_1\int _{}dx^{\prime}_2 \int
_{}dy^{\prime}_1\int _{}dy^{\prime}_2 \;\: W[\lambda](x^{\prime}_1,
y^{\prime}_1;x^{\prime}_2,y^{\prime}_2)\:
\prod_{j=1}^2 G_\tau(x_j|x^{\prime}_j) G_\tau(y_j|y^{\prime}_j) \:
\label{conv}
\end{eqnarray}
where $W[\lambda]$ is initial Wigner function of
the TWB, and the Green functions $G_\tau(x_j|x^{\prime}_j)$ are
given by
\begin{eqnarray}
G_\tau(x_j|x^{\prime}_j)=\frac{1}{\sqrt{2\pi
D^2}}\exp\left[-\frac{ (x_j-x^{\prime}_je^{-\frac12 \gamma
\tau})^2} {2 D^2}\right] \;,\quad D^2=\frac{1}{4\gamma
}(1-e^{-\gamma \tau}) \;. \label{green}
\end{eqnarray}
\end{widetext}
The Wigner function $W_\tau$ can be obtained by the convolution
(\ref{conv}), which can be easily evaluated since the initial
Wigner function is Gaussian. The form of $W_\tau$ is the same of
$W[\lambda]$ with the variances changed to
\begin{eqnarray}
\sigma_+^2 \longrightarrow
e^{\gamma\tau}\left(\sigma_+^2+D^2\right)\qquad \sigma_-^2
\longrightarrow e^{\gamma\tau}\left(\sigma_-^2+D^2\right)
   \label{evolvedVar}\:.
\end{eqnarray}
Inserting the Wigner functions of the POVM $\Pi_{\alpha\eta}$ and
of the evolved TWB in Eq. (\ref{rhgenW}) we obtain the teleported
state in the general case. This still has the form (\ref{tel}),
however with the parameter $K$ now given by
\begin{eqnarray}
K = K_0 e^{\Gamma t} + (2M+1)(e^{\Gamma
t}-1) + D_\eta^2 \label{kappar}\;.
\end{eqnarray}
Eqs. (\ref{tel}) and (\ref{kappar}) summarize the possible
effects that degrade the quality of teleportation. In the special
case of coherent state teleportation one has $S=|z\rangle\langle z|$,
which corresponds to original optical CV teleportation
experiments \cite{kim}. The fidelity $F=\langle z|\sigma
|z\rangle$ can be evaluated straightforwardly as the overlap of
the Wigner functions. Since $W[z](\alpha)=\frac2\pi
\: e^{-2|\alpha -z|^2}$ is the Wigner function of a coherent state,
we have $$ F = \frac{1}{1+K_0 e^{\Gamma t}+(e^{\Gamma t-1})(2M+1)+D^2_\eta}\:.$$
In order to verify quantum teleportation, {\em i.e.} to show that
the scheme is a truly nonlocal protocol, the fidelity should fulfill
the bound $F>1/2$ \cite{kim}, {\em i.e.}
$$ K_0 e^{\Gamma t}+(e^{\Gamma t-1})(2M+1)+D_\eta^2 <1\:. $$
Therefore, given the value of the parameters $\Gamma$, $M$, and $\eta$,
in order to verify quantum teleportation, one should use TWB with a number of
photons satisfying the bound
\begin{eqnarray}
1+N-\sqrt{N(N+2)} < e^{-\Gamma t} \left[1-D^2_\eta -(2M+1)(e^{\Gamma
t}-1)\right]
\label{boundN}\;.
\end{eqnarray}
\section{Summary and conclusions}
\label{s:outro}
A measurement performed on one beam of a TWB reduces the other one
according to the projection postulate. This effect is an
intrinsic quantum mechanism to achieve effective nonlinear dynamics. We have
analyzed in details the use of conditional measurement on TWB to
generate and manipulate quantum states of light. In particular, we
have studied realistic measurement schemes taking into account
imperfections of the apparatuses, such as detection quantum
efficiency and finite resolution. \par The first kind of
measurement we have analyzed is {\sc on/off} photodetection which
provides a reliable scheme to check nonclassical light. The
nonclassicality and its verification are robust against the TWB energy
and the detector efficiency. The second apparatus is homodyne
detector, whose action on TWB represents a tunable source of
squeezed light, with high conditional probability and robustness
to experimental imperfections. In particular, in the ideal case,
the conditional output state is a pure minimum uncertainty state
which two features: the amount of squeezing is independent on the
outcome of the measurement, which only influences the coherent
amplitude, and the most probable conditional state is a squeezed
vacuum. Taking into account the effect of nonunit quantum efficiency 
and finite resolution, we have that the conditional state is no longer a pure
state, however, still showing squeezing for quantum efficiency
larger than $\eta > 1/2$ and for a large range of the homodyne
outcomes.
\par
Finally, we have shown how to describe optical CV teleportation as
a conditional measurement of the sum- and difference-quadratures
of two modes. We found that realistic CV teleportation with finite
amount of entanglement is equivalent to a Gaussian channel with
$K_0\simeq (2N)^{-1}$ background photons applied to the input
state. Using Wigner functions we have also shown that the
teleportation in the general case, {\em i.e.} taking into account
the degrading effects of finite amount of entanglement,
decoherence due to losses, and imperfect detection, still
corresponds to a Gaussian channel, however with an increased number 
of background photons [see Eq. (\ref{kappar})]. A
bound on the average TWB energy, in order to verify quantum teleportation, 
has been derived. \par  We conclude that performing conditional
measurements on entangled twin-beam is a powerful and robust
method to engineering nonclassical states of light.
\section*{Acknowledgments}
This work has been sponsored by the INFM through the project PRA-2002-CLON, by
MIUR through the PRIN project {\em Decoherence control in quantum information
processing} and by EEC through the project IST-2000-29681 (ATESIT).
MGAP is research fellow at Collegio Alessandro Volta, Pavia.

\end{document}